\documentstyle[preprint,eqsecnum,aps]{revtex} 
\begin{document}



\def\YM/{Yang\discretionary{-}{}{-}Mills}
\def\FT/{Freedman\discretionary{-}{}{-}Townsend}

\def\pde/{partial differential equation}

\hyphenation{
all
along
anti
ap-pen-dix 
co-tan-gent
equa-tion
equa-tion-s
equiv-a-lent
evo-lu-tion
fields
form
iden-ti-ty
iden-ti-ties
im-por-tant
its
La-grang-ian
La-grang-ian-s
next
nev-er
prod-uct
real
sca-lar
sym-me-try
sym-me-tries
tak-en
tan-gent
term
two
use
use-s
vari-a-tion
vari-a-tion-s
}


\def\Rnum{{\bf R}}
\def\Cnum{{\bf C}}
\def\one{{\bf 1}}
\def\vec#1{{\bf #1}}
\def\su#1{$SU(#1)$}
\def\so#1{$SO(#1)$}

\def\eqref#1{(\ref{#1})}
\def\eqrefs#1#2{(\ref{#1}) and~(\ref{#2})}
\def\eqsref#1#2{(\ref{#1}) to~(\ref{#2})}

\def\Eqref#1{Eq.~(\ref{#1})}
\def\Eqrefs#1#2{Eqs.~(\ref{#1}) and~(\ref{#2})}
\def\Eqsref#1#2{Eqs.~(\ref{#1}) to~(\ref{#2})}

\def\secref#1{Sec.~\ref{#1}}
\def\secrefs#1#2{Sec.~\ref{#1} and~\ref{#2}}

\def\appref#1{App.~\ref{#1}}

\def\citeref#1{Ref.~\cite{#1}}

\def\Cite#1{${\mathstrut}^{\cite{#1}}$}

\def\tableref#1{Table~\ref{#1}}

\def\figref#1{Fig.~\ref{#1}}

\hyphenation{Eq Eqs Sec App Ref Fig}

\def\EQ{\begin{equation}}
\def\EQs{\begin{eqnarray}}
\def\endEQ{\end{equation}}
\def\endEQs{\end{eqnarray}}

\def\proclaim#1{\medbreak
\noindent{\it {#1}}\par\medbreak}
\def\Proclaim#1#2{\medbreak
\noindent{\bf {#1}}{\it {#2}}\par\medbreak}


\def\newline{\hfil\break}
\def\fewquad{\qquad\qquad}
\def\severalquad{\qquad\fewquad}
\def\manyquad{\qquad\severalquad}
\def\manymanyquad{\manyquad\manyquad}

\def\downupindices#1#2{{}^{}_{#1}{}_{}^{#2}}
\def\updownindices#1#2{{\mathstrut}_{}^{#1}{}^{}_{#2}}
\def\mixedindices#1#2{{\mathstrut}^{#1}_{#2}}
\def\downindex#1{{\mathstrut}^{\mathstrut}_{#1}}
\def\upindex#1{{\mathstrut}_{\mathstrut}^{#1}}

\def\tensor#1#2#3{{#1}\mixedindices{#2}{#3}}
\def\covector#1#2{{#1}\downindex{#2}}
\def\vector#1#2{{#1}\upindex{#2}}

\def\id#1#2{\delta\downupindices{#1}{#2}}
\def\cross#1#2{\epsilon\,\downupindices{#1}{#2}}
\def\vol#1{\epsilon\,\downindex{#1}}
\def\invvol#1{\epsilon\,\upindex{#1}}

\def\metric#1{\eta\downindex{#1}}
\def\invmetric#1{\eta\upindex{#1}}

\def\measure#1{d\,x^{#1}}

\def\A#1#2{\tensor{A}{#1}{#2}}
\def\F#1#2{\tensor{F}{#1}{#2}}
\def\stF#1#2{\tensor{*F}{#1}{#2}}
\def\X#1{\vector{\xi}{#1}}
\def\Xsub#1#2{\tensor{\xi}{#1}{#2}}
\def\Y#1#2{\tensor{Y}{#1}{#2}}
\def\Yinv#1#2{\tensor{Y^{-1}}{#1}{#2}}
\def\YinvF#1#2{\tensor{(Y^{-1}*F)}{#1}{#2}}
\def\K#1#2{\tensor{K}{#1}{#2}}
\def\YinvE#1#2{\tensor{(Y^{-1}E)}{#1}{#2}}
\def\E#1#2{\tensor{E}{#1}{#2}}
\def\J#1#2{\tensor{J}{#1}{#2}}
\def\Q#1{\vector{Q}{#1}}
\def\T#1{T\downindex{#1}}

\def\B#1#2{{\rm B}\downupindices{#1}{#2}}
\def\duB#1#2{{\rm B}\updownindices{#1}{#2}}
\def\e#1{{\bf e}\downindex{#1}}
\def\v#1{v\downindex{#1}}
\def\vup#1{v\upindex{#1}}
\def\u#1{u\downindex{#1}}
\def\uup#1{u\upindex{#1}}
\def\x#1{x\downindex{#1}}
\def\xup#1{x\upindex{#1}}
\def\y#1{y\downindex{#1}}
\def\yup#1{y\upindex{#1}}
\def\w#1#2{\tensor{w}{#1}{#2}}
\def\H#1{H\downindex{#1}}
\def\kv#1{\vector{\zeta}{#1}}
\def\C#1#2{{\rm C}\downupindices{#1}{#2}}
\def\k#1#2{k\downupindices{#1}{#2}}
\def\V#1{V\downindex{#1}}
\def\vecA#1{\vec{A}\downindex{#1}}
\def\vecF#1{\vec{F}\downindex{#1}}
\def\vecK#1{\vec{K}\downindex{#1}}
\def\vecX{\vec{\X{}}}
\def\vecunitX{\vec{\hat\X{}}}
\def\vecA#1{\vec{A}\downindex{#1}}
\def\ad#1{{\rm ad}\downindex{#1}}
\def\adinv#1{{\rm ad}\mixedindices{-1}{#1}}
\def\proj#1{{\rm P}\downindex{\perp #1}}
\def\Hsp{\vec{H}}
\def\Hperp{\vec{H_\perp}}
\def\Asp{\vec{A}}
\def\Nsp{\vec{N}}

\def\der#1{\partial\downindex{#1}}
\def\duder#1{\partial\upindex{#1}}

\def\Der#1{D\downindex{#1}}
\def\duDer#1{D\upindex{#1}}

\def\Lie#1{{\cal L}{}_{#1}}

\def\frac#1#2{{\textstyle {#1 \over #2}}}

\def\Rnum{\bold R}
\def\Cnum{\bold C}

\preprint{ Published in J. Math. Phys., 38 (1997) p3399-3413}

\title{ 
Novel generalization of three-dimensional Yang-Mills theory}

\author{Stephen C. Anco\cite{email}}
\address{
Department of Mathematics\\ 
University of British Columbia, 
Vancouver, BC Canada V6T 1Z1 \\
and Department of Physics\\
University of Maryland,
College Park, MD 20742
}

\maketitle

\begin{abstract}
A class of new nonabelian gauge theories 
for vector fields on three manifolds is presented.
The theories describe a generalization of three-dimensional \YM/ theory
featuring a novel nonlinear gauge symmetry and field equations
for Lie-algebra valued vector potential fields.
The nonlinear form of the gauge symmetry and field equations
relies on the vector cross-product and vector curl operator
available only in three dimensions
and makes use of an auxiliary Lie bracket 
together with the Lie bracket used in \YM/ theory.
A gauge covariant formulation of the new theories is given
which utilizes the covariant derivative and curvature
from the geometrical formulation of \YM/ theory.
Further features of the new theories are discussed.
\end{abstract}

\narrowtext\newpage

\section{Introduction}
\label{intro}

The wide ranging importance of 
gauge theories of vector fields in mathematics and physics 
raises interest in finding new types of these theories,
especially generalizations of \YM/ theory.
Recently, a new gauge theory \cite{abelianth} 
of this type has been constructed 
for vector fields on three-dimensional spacetimes.
The theory can be understood to describe 
a nonlinear generalization of abelian \YM/ theory
with the novel feature of a nonlinear gauge symmetry 
for abelian vector potential fields.
The form of the gauge symmetry and field equations
relies on vector cross-product and curl operations,
limiting the generalization as a gauge theory of vector fields
to work in three dimensions only. 

This generalization is closely analogous in structure 
to the \FT/ \cite{FTth} 
nonlinear generalization of abelian gauge theory 
for antisymmetric tensor fields on four-dimensional spacetimes.
In particular, under the replacement of 
three-dimensional vector fields by
d-dimensional antisymmetric tensor fields of rank d-2,
the generalization of abelian \YM/ theory can be extended naturally
as a gauge theory of antisymmetric tensor fields 
from three dimensions to d dimensions,
where the four-dimensional theory is equivalent to 
the theory constructed by Freedman and Townsend. 
The structure of these theories as \YM/ theory generalizations 
suggests some extensions 
that can be pursued to give additional gauge theories of 
vector fields and antisymmetric tensor fields. 

This paper is addressed to extending 
the \YM/ theory generalization in three dimensions 
from the abelian case to the nonabelian case, 
giving a class of new nonabelian gauge theories of 
vector fields on three-dimensional spacetimes,
which was announced in \citeref{abelianth}. 
The \su2 case of this generalization is worked out in \secref{newtheory}
and results in a novel \su2 theory 
featuring a new form of nonlinear gauge symmetry and field equations.
The construction of the theory relies on 
vector cross-product and curl operations available in three dimensions,
and also uses the \su2 Lie algebra 
along with a related auxiliary Lie algebra. 
In contrast to the abelian case, 
where the analogous auxiliary Lie algebra is allowed to be arbitrary,
the auxiliary Lie algebra in the \su2 case 
is fixed through an algebraic condition required by 
invariance of the action principle for the field equations 
under the gauge symmetry. 
As a consequence, in the \su2 case, 
the nonlinear structure of the generalization is unique. 

Extension of the \su2 case 
to the general nonabelian case of the generalization 
is considered in \secref{extension},
which leads to a general class of nonabelian theories
with features similar to the \su2 theory. 
In \secref{discuss} 
some additional features and extensions of the generalization 
are discussed.

\section{New SU(2) gauge theory}
\label{newtheory}

\subsection{Formulation}

We start from the vector potential and field strength
of \su2 \YM/ theory on a three-dimensional manifold. 
The structure we require on the manifold \cite{global}
is a metric $\metric{\mu\nu}$ 
and a volume form $\vol{\mu\nu\sigma}$ 
normalized with respect to metric,
along with a compatible derivative operator $\der{\mu}$ 
determined by the metric. 
We also use the structure of the Lie algebra of \su2
in an arbitrary fixed basis $\{\e{a}\}_{a=1,2,3}$,
with Lie algebra multiplication $[\e{a},\e{b}]=\cross{ab}{c} \e{c}$
and Killing metric $(\e{a},\e{b})=\id{ab}{}$.
We fix the Killing metric components $\id{ab}{}$ in terms of 
the multiplication structure constants $\cross{ab}{c}$ by 
$\cross{ad}{e} \cross{be}{d} = -2\id{ab}{}$,
and we denote the components of the inverse of the Killing metric by 
$\id{}{ab}$,
with $\id{ab}{} \id{}{bc} = \id{a}{c}$
denoting the components of the identity operator in the \su2 Lie algebra.
In addition, we denote the inverse of $\metric{\mu\nu}$ by
$\invmetric{\mu\nu}$,
where $\metric{\mu\nu} \invmetric{\nu\sigma} = \id{\mu}{\sigma}$
is the identity tensor on the manifold.
(Throughout, latin letters are used for internal Lie algebra indices,
and greek letters are used for manifold coordinate indices,
with all indices running from $1$ to $3$.)

We represent the vector potential in the \su2 basis 
by a set of cotangent vector fields 
$\{\A{a}{\mu}\}_{a=1,2,3}$
and similarly represent the field strength of the vector potential 
by a set of antisymmetric tensor fields
$\{\F{a}{\mu\nu}\}_{a=1,2,3}$.
The field strength tensors $\F{a}{\mu\nu}$ 
are given in terms of the fields $\A{a}{\mu}$ by
\EQ\label{ymstrength}
\F{a}{\mu\nu} = 
\der{[\mu}\A{a}{\nu]} + \case{1}{2} \cross{bc}{a} \A{b}{\mu}\A{c}{\nu}
\endEQ
The dual of these tensors are vectors 
$\F{a}{\sigma} = \cross{\sigma}{\mu\nu} \F{a}{\mu\nu}$
which are constructable entirely by 
curl and cross-product operations 
along with \su2 multiplication,
\EQ\label{ymdualF}
\F{a}{\sigma} = 
\cross{\sigma}{\mu\nu}\der{\mu}\A{a}{\nu} 
+ \case{1}{2} \cross{bc}{a} \cross{\sigma}{\mu\nu} \A{b}{\mu}\A{c}{\nu}
\endEQ
where 
$\cross{\sigma}{\mu\nu} 
= \vol{\sigma\alpha\beta} \invmetric{\mu\alpha} \invmetric{\nu\beta}$
is the cross-product tensor
and $\cross{\sigma}{\mu\nu}\der{\mu}$ is the curl operator.

To proceed with the construction of the new \su2 gauge theory,
we introduce structure constants $\B{ab}{c}$ 
defining an additional Lie algebra multiplication 
of the \su2 basis vectors $\e{a}$. 
This multiplication will later be fixed 
through an algebraic condition imposed by gauge invariance.
Using the structure constants,
we now construct the field tensors
\EQ\label{Y}
\Y{d \nu}{\tau b} = 
\id{\tau}{\nu} \id{b}{d} 
- \cross{\tau}{\nu\sigma} \duB{d}{be} \A{e}{\sigma}
\endEQ
linearly in terms of $\A{e}{\sigma}$,
where 
\EQ
\duB{d}{be}= \id{}{da} \id{ce}{} \B{ab}{c}
\endEQ
The tensors $\Y{d \nu}{\tau b}$ represent 
components of a linear map $Y$ on Lie-algebra valued cotangent vectors. 
We define tensors $\Yinv{a \tau}{\mu d}$ to represent 
components of the inverse linear map $\Yinv{}{}$, such that
\EQ\label{Yinv}
\Yinv{a \tau}{\mu d} \Y{d \nu}{\tau b} = 
\id{b}{a} \id{\mu}{\nu} = 
\Y{a \tau}{\mu d} \Yinv{d \nu}{\tau b}
\endEQ
with $\A{e}{\sigma}$ restricted by the condition 
$det(Y)\neq 0$ necessary for invertibility of $Y$. 
We also define the associated tensors
\EQs
&& \Y{\mu\nu}{ab} 
= \id{ad}{} \invmetric{\mu\tau} \Y{d \nu}{\tau b}
= \id{ab}{} \invmetric{\mu\nu}
- \cross{\sigma}{\mu\nu} \B{ab}{c} \id{ce}{} \A{e}{\sigma}
\label{Ytensor}\\
&& \Yinv{\sigma\tau}{cd} 
= \id{ca}{} \invmetric{\sigma\mu} \Yinv{a \tau}{\mu d}
\label{Yinvtensor}
\endEQs
which represent components of 
nondegenerate bilinear forms on Lie-algebra valued cotangent vectors.
The bilinear forms are symmetric 
$\Y{\mu\nu}{ab}  = \Y{\nu\mu}{ba}$ 
and $\Yinv{\sigma\tau}{cd} = \Yinv{\tau\sigma}{dc}$
due to the antisymmetry 
$\cross{\sigma}{\mu\nu} = \cross{\sigma}{[\mu\nu]}$
and  $\B{ab}{c} = \B{[ab]}{c}$.

From the previous structure 
the theory is constructed as follows. 
Field strength vectors in the theory are defined by 
\EQ\label{K}
\K{a}{\mu} = \Yinv{a \nu}{\mu b} \F{b}{\nu}
\endEQ
In terms of these field strengths 
the Lagrangian of the theory is given quadratically by 
\EQ\label{L}
L= \Yinv{\sigma\tau}{cd} \F{c}{\sigma} \F{d}{\tau}
= \Y{\mu\nu}{ab} \K{a}{\mu} \K{b}{\nu}
\endEQ
The gauge symmetry of the theory is given by 
the infinitesimal transformations
\EQ\label{gaugesymm}
\delta\A{a}{\mu} = 
\der{\mu} \X{a} + ( \cross{bc}{a} \A{b}{\mu} 
+ \duB{a}{dc} \K{d}{\mu} ) \X{c}
\endEQ
involving a set of arbitrary functions $\X{a}$.

Gauge invariance of the theory requires $L$ to vary 
into a complete divergence $\der{\nu} S^\nu$ 
under an arbitrary gauge transformation $\delta\A{a}{\mu}$,
where $S^\nu$ is some local function of 
$\A{a}{\mu}$, $\X{a}$, and their derivatives. 
This requirement now leads to the algebraic condition
fixing the structure constants $\B{ab}{c}$.

Consider an arbitrary infinitesimal variation of $L$,
\EQ\label{varL}
\delta L= \delta( \Yinv{\mu\nu}{ab} \F{a}{\mu} \F{b}{\nu} )
\endEQ
Varying $\F{a}{\mu}$ contributes the terms 
\EQs\label{varF}
2 \Yinv{\mu\nu}{ab} \delta \F{a}{\mu} \F{b}{\nu} 
&& = 2 \invmetric{\sigma\mu} \id{ca}{} \K{c}{\sigma} \delta\F{a}{\mu}
\nonumber\\
&& = 2 \K{c}{\sigma} \invvol{\sigma\nu\tau} \id{ca}{}
( \der{\nu}\delta\A{a}{\tau} + \cross{de}{a} \A{d}{\nu} \delta\A{e}{\tau} )
\endEQs
and varying $\Yinv{\mu\nu}{ab}$ contributes the terms
\EQs\label{varYinv}
\delta\Yinv{\mu\nu}{ab} \F{a}{\mu} \F{b}{\nu} 
&& = - \F{a}{\mu} \F{b}{\nu} \Yinv{c \mu}{\sigma a} \Yinv{d \nu}{\tau b}
\delta \Y{\sigma\tau}{cd}
\nonumber\\
&& = \K{c}{\sigma} \K{d}{\tau} 
( \invvol{\sigma\tau\nu} \B{cde}{} \delta\A{e}{\nu} )
\endEQs
through use of \Eqsref{Yinv}{Yinvtensor},
with $\invvol{\sigma\tau\nu}
= \invmetric{\sigma\mu}\cross{\mu}{\tau\nu}$
and $\B{cde}{} = \B{cd}{b} \id{be}{}$.
We now substitute the gauge symmetry \eqref{gaugesymm} 
for the variations $\delta\A{\mu}{a}$
and group the terms into a quadratic expression 
in powers of $\B{ab}{c}$ and $\cross{ab}{c}$.

The linear part involving $\B{ab}{c}$ in $\delta L$ 
consists of the terms 
\EQs
&& 2\invvol{\nu\sigma\tau} \B{bde}{} \K{b}{\nu} \der{\sigma}
( \K{d}{\tau}\X{e} ) 
+ \invvol{\nu\tau\sigma} \B{bde}{} \K{b}{\nu} \K{d}{\tau} \der{\sigma} \X{e}
\nonumber\\&&
= \invvol{\nu\sigma\tau} \B{bde}{} \der{\sigma} 
( \X{e} \K{b}{\nu} \K{d}{\tau} )
\endEQs
which combine to yield a complete divergence,
using the antisymmetry property of Lie algebra multiplication,
$\B{bd}{a} = \B{[bd]}{a}$.
The quadratic part involving $\B{ab}{c}$ in $\delta L$ 
is given by one term
\EQ\label{BBterm}
\B{bc}{e} \B{edn}{} \X{n} \invvol{\nu\sigma\tau} 
\K{b}{\nu} \K{c}{\sigma} \K{d}{\tau}
\endEQ
The antisymmetry $\invvol{\nu\sigma\tau}=\invvol{[\nu\sigma\tau]}$ 
implies the product $\B{bc}{e}\B{edn}{}$ 
is antisymmetric in its indices $bcd$
and thus vanishes by the Jacobi property of Lie algebra multiplication,
$\B{[bc}{e} \B{|e|d]}{a} =0$.
Consequently, the term \eqref{BBterm} vanishes.
Next, turning to the terms from the linear and quadratic parts
involving $\cross{ab}{c}$ in $\delta L$ 
yields
\EQs
&& 2\invvol{\nu\sigma\tau} \K{c}{\nu}  
( \cross{dec}{} \der{\sigma} ( \A{d}{\tau} \X{e} ) 
+ \cross{dec}{} \A{e}{\tau} \der{\sigma} \X{d} 
+ \cross{dec}{} \cross{mn}{d} \A{e}{\tau} \A{m}{\sigma} \X{n} )
\nonumber\\&& 
= 2\invvol{\nu\sigma\tau} \K{c}{\nu} 
( \cross{dec}{} \der{\sigma} \A{d}{\tau} \X{e} + 
\frac{1}{2} \cross{dnc}{}\cross{em}{d} \A{e}{\tau} \A{m}{\sigma} \X{n} )
\nonumber\\&& 
= 2\cross{dec}{} \invmetric{\nu\mu} 
\K{c}{\nu} \F{d}{\mu} \X{e}
\label{KAterms}
\endEQs
after we have rearranged 
$2\cross{d[e|c|}{}\cross{m]n}{d}= \cross{dnc}{}\cross{em}{d}$
by the Jacobi property of \su2 multiplication,
where $\cross{dec}{}=\cross{de}{b} \id{bc}{}$.
Using the relations \eqrefs{Y}{K}, we express
\EQ\label{Fid}
\F{d}{\mu}= \Y{d \sigma}{\mu a } \K{a}{\sigma} 
= \K{d}{\mu} -\cross{\mu}{\sigma\tau}\duB{d}{am} \K{a}{\sigma}\A{m}{\tau}
\endEQ
in the term \eqref{KAterms},
which leads to 
\EQ\label{KFterm}
2\cross{dec}{} \invmetric{\nu\mu} 
\K{c}{\nu} \F{d}{\mu} \X{e} 
= 2\cross{cde}{} \invmetric{\nu\mu} 
\K{c}{\nu} \K{d}{\mu} \X{e} 
-2\cross{ec}{d} \B{dam}{} \invvol{\nu\sigma\tau} 
\K{c}{\nu} \K{a}{\sigma} \A{m}{\tau} \X{e}
\endEQ
by the \su2 multiplication relation 
$\cross{dec}{}= \cross{cde}{}= \cross{ecd}{}$.
The antisymmetry $\cross{cde}{}=\cross{[cd]e}{}$ 
implies the term involving 
$2\cross{cde}{} \invmetric{\nu\mu} \K{c}{\nu} \K{d}{\mu}$ 
in \Eqref{KFterm} vanishes. 
Finally, the other term from \Eqref{KFterm}
can be combined with all the terms in $\delta L$
from the remaining quadratic parts, 
\EQs
&&
-2\invvol{\nu\sigma\tau} \B{dam}{} \cross{ec}{d} 
\K{c}{\nu} \K{a}{\sigma} \A{m}{\tau} \X{e}
+ 2\invvol{\nu\sigma\tau} \B{dmn}{} \cross{eb}{d} 
\K{b}{\nu} \K{m}{\sigma} \A{e}{\tau} \X{n}
+ \invvol{\nu\sigma\tau} \B{bce}{} \cross{mn}{e} 
\K{b}{\nu} \K{c}{\sigma} \A{m}{\tau} \X{n}
\nonumber\\&& 
=\invvol{\nu\sigma\tau} \K{b}{\nu} \K{c}{\sigma} \A{a}{\tau} \X{d} \H{bcad}
\endEQs
where we have defined
\EQ\label{H}
\H{bcad}= 
2\cross{a[b}{e} \B{|e|c]d}{} 
- 2\cross{d[b}{e} \B{|e|c]a}{} + \cross{ad}{e} \B{bce}{}
\endEQ

Assembling the previous terms in the gauge symmetry variation $\delta L$ 
now yields
\EQ
\delta L = \der{\sigma} ( \invvol{\sigma\tau\nu} \B{bde}{} 
\K{b}{\nu} \K{d}{\tau} \X{e} )
+ \invvol{\nu\sigma\tau} \K{b}{\nu} \K{c}{\sigma} \A{a}{\tau} \X{d} \H{bcad}
\endEQ
which is required to equal a complete divergence.
Because the second term involves no derivatives of $\X{d}$,
it is not a divergence and therefore must vanish.
This implies
\EQ\label{condition}
0=\H{bcad}
\endEQ
constituting an algebraic condition on the structure constants $\B{ab}{c}$.

We now solve \Eqref{condition} 
to determine $\B{ab}{c}$ subject to the structure constant properties
\EQ
\B{ab}{c} = \B{[ab]}{c} , 
\B{[ab}{d} \B{c]d}{e} =0
\endEQ
We contract \Eqref{condition} with $\cross{n}{ad}$ 
and use the \su2 identities
\EQs
&& \cross{ad}{e} \cross{n}{ad} = 2\id{n}{e}
\label{firstid}\\
&& \cross{ab}{e} \cross{n}{ad} = \id{b}{d} \id{n}{e} - \id{bn}{} \id{}{de}
\label{secondid}
\endEQs
where $\cross{n}{ad}$ is the transpose of $\cross{ad}{n}$ in the \su2 metric. 
This yields, using the property $\B{ab}{c} = \B{[ab]}{c}$,
\EQ\label{contractedH}
0=\cross{n}{ad} \H{bcad} 
= 4\id{n[b}{}\B{c]ad}{}\id{}{ad} + 2\B{ncb}{} -2\B{nbc}{} + 2\B{bcn}{}
\endEQ
Permuting the free indices in \Eqref{contractedH} 
and adding the resulting equations now leads to 
\EQ
0=\id{n[b}{} \B{c]ad}{} \id{}{ad} + \B{bcn}{}
\endEQ
This algebraic relation implies
\EQ\label{sutwoB}
\B{bcn}{} = \id{nb}{} \v{c} - \id{nc}{} \v{b}
\endEQ
for some arbitrary constants $\v{c}$.
The expression \eqref{sutwoB} is easily checked to satisfy \Eqref{condition}
as well as the required properties
\EQs
&& \B{bcn}{} = 2\id{n[b}{} \v{c]} = \B{[bc]n}{} , 
\\&& \B{[bc}{n} \B{d]n}{a} 
= 2 \id{[b}{n} \v{c} \id{d]}{a} \v{n} - 2 \id{[b}{n} \v{c} \v{d]} \id{n}{a} 
= 4 \v{[b} \v{c} \id{d]}{a} =0
\label{sutwoproperty}
\endEQs

Thus, from gauge invariance, 
the algebraic condition \eqref{condition} 
fixes the auxiliary structure constants \eqref{sutwoB}
which appear in the field strength \eqref{K},
the Lagrangian \eqref{L}, and the gauge symmetry \eqref{gaugesymm}
of the theory.

These structure constants define an auxiliary Lie algebra multiplication
\EQ\label{basismult}
[\e{a},\e{b}]_B = \B{ab}{c} \e{c} = 2\v{[b}\e{a]}
\endEQ
in terms of the \su2 basis vectors $\e{a}$. 
The auxiliary multiplication is related to \su2 multiplication,
\EQ\label{auxmult}
[\e{a},\e{b}]_B = [[\e{a},\e{b}],\vec{v}]
\endEQ
using the Lie-algebra vector $\vec{v} = \v{b} \id{}{ba} \e{a}$ 
determined by the constants $\v{b}$. 
This relationship \eqref{auxmult} directly follows from
expressing the auxiliary structure constants 
in terms of the \su2 structure constants
\EQ\label{Bid}
\cross{ab}{e}\cross{ed}{c} \id{}{dn} \v{n} 
= 2\id{[a}{n}\id{b]}{c} \v{n}
=2 \v{[a}\id{b]}{c} = \B{ab}{c}
\endEQ
with the use of the \su2 identity \eqref{secondid}
and some index rearrangements.

The structure of the auxiliary multiplication can be understood as follows.
Fix two linearly independent Lie-algebra vectors 
$\vec{w}{}_1=\w{a}{1} \e{a}$ and $\vec{w}{}_2=\w{a}{2} \e{a}$ 
such that $\w{a}{1}\v{a}=0=\w{a}{2}\v{a}$,
so the vectors $\{\vec{w}{}_1,\vec{w}{}_2,\vec{v}\}$ 
provide a Lie algebra basis
with $\vec{v}$ orthogonal to $\vec{w}{}_1$ and $\vec{w}{}_2$ 
in the \su2 Killing metric. 
From \Eqref{basismult} 
the auxiliary multiplication of this basis is given by 
\EQs
&& [\vec{w}{}_1,\vec{w}{}_2]_B 
= 2 \w{a}{1} \w{b}{2} \v{[b}\e{a]} =0
\\&& [\vec{w}{}_1,\vec{v}]_B 
= 2 \w{a}{1} \id{}{cb} \v{c} \v{[b}\e{a]} = \vec{w}{}_1 |\vec{v}|^2
\\&& [\vec{w}{}_2,\vec{v}]_B 
= 2 \w{a}{2} \id{}{cb} \v{c} \v{[b}\e{a]} = \vec{w}{}_2 |\vec{v}|^2
\endEQs
where $|\vec{v}|^2=\id{}{ab}\v{a}\v{b}$ is the \su2 norm squared of $\vec{v}$.
This multiplication structure represents 
a three-dimensional nilpotent Lie algebra \cite{isomorphic}
which is the semi-direct product of 
the one-dimensional Lie algebra spanned by $\vec{v}$
with the two-dimensional abelian Lie algebra 
spanned by $\vec{w}{}_1$ and $\vec{w}{}_2$,
where $\case{1}{|\vec{v}|^2}\vec{v}$ acts by 
multiplication as an identity element on $\vec{w}{}_1$ and $\vec{w}{}_2$.

\subsection{Features}

The variation of the Lagrangian \eqref{L} 
as assembled from \Eqsref{varL}{varYinv}
yields the field equations for $\A{a}{\mu}$,
\EQ\label{fieldeqs}
\E{a}{\mu} = 2\cross{\mu}{\sigma\nu} (
\der{\sigma} \K{a}{\nu} + \cross{bc}{a} \A{b}{\sigma} \K{c}{\nu}
+ \case{1}{2} \B{bc}{a}\K{b}{\sigma} \K{c}{\nu} ) 
=0
\endEQ
The fields $\A{a}{\mu}$ appear in $\E{a}{\mu}$ nonpolynomially
through the field strengths $\K{a}{\mu}$,
producing a novel form of nonlinear coupling of the fields,
which is controlled by the auxiliary structure constants $\B{bc}{a}$
together with the \su2 structure constants $\cross{bc}{a}$.

The field strengths have two basic properties.
First, the \su2 Bianchi identity 
$0=\duder{\mu} \F{a}{\mu} 
+ \cross{bc}{a} \A{b}{\mu} \F{c}{\nu} \invmetric{\mu\nu}$
leads to the differential identity on $\K{a}{\mu}$,
\EQ\label{Keqs}
\duder{\mu} \K{a}{\mu} 
+ ( \cross{bc}{a} \A{b}{\mu} \K{c}{\nu} 
+ \duB{a}{dc} \K{d}{\mu} \K{c}{\nu} )\invmetric{\mu\nu}
= \case{1}{2} \duB{a}{dc} \E{d}{\mu} \A{c}{\nu} \invmetric{\mu\nu}
\endEQ
where $\duder{\mu} = \invmetric{\mu\nu} \der{\nu}$.
Consequently, for solutions of the field equations,
$\K{a}{\mu}$ satisfies \su2 field equations of the form 
\EQs
&& \invmetric{\mu\nu} ( \der{\mu}\K{a}{\nu} 
+ \cross{bc}{a} \A{b}{\mu}\K{c}{\nu} )
= -\duB{a}{dc} \K{d}{\mu} \K{c}{\nu} \invmetric{\mu\nu}
\\&&
\cross{\mu}{\sigma\nu} ( \der{\sigma} \K{a}{\nu} 
+ \cross{bc}{a} \A{b}{\sigma} \K{c}{\nu} )
= -\case{1}{2} \B{bc}{a}\K{b}{\sigma} \K{c}{\nu} \cross{\mu}{\sigma\nu} 
\endEQs
involving scalar and vector source terms 
generated quadratically from $\K{a}{\mu}$.
Second, under the gauge symmetry \eqref{gaugesymm} on $\A{a}{\mu}$,
the field strengths $\K{a}{\mu}$ have the transformation
\EQ\label{Kgaugesymm}
\delta\K{a}{\mu} =
\cross{bc}{a} \K{b}{\mu} \X{c} + \case{1}{2} \YinvE{a}{e \mu} \X{e}
\endEQ
where $\YinvE{a}{e \mu} = \Yinv{a \nu}{\mu b} \E{c}{\nu} \duB{b}{ce}$.
For solutions of the field equations,
the form of \Eqref{Kgaugesymm} 
represents an infinitesimal rotation of $\K{a}{\mu}$ as an \su2 vector,
and so \su2 invariants constructed from $\K{a}{\mu}$ 
yield gauge symmetry invariants in the theory.

The gauge symmetry of the theory has a closed commutator structure 
on solutions of the field equations.
From \Eqrefs{gaugesymm}{Kgaugesymm},
calculating the commutator of two infinitesimal gauge transformations 
$\delta_1\A{a}{\mu}$ and $\delta_2\A{a}{\mu}$ 
involving sets of arbitrary functions $\Xsub{a}{1}$ and $\Xsub{a}{2}$
yields
\EQ\label{commutator}
[\delta_1,\delta_2]\A{a}{\mu} = 
\delta_3\A{a}{\mu} + \duB{a}{c[d} \YinvE{c}{e]\mu} \Xsub{e}{2} \Xsub{d}{1}
\endEQ
where the infinitesimal gauge transformation $\delta_3\A{a}{\mu}$ 
involves the set of functions 
$\Xsub{c}{3} = \cross{ed}{c} \Xsub{e}{2} \Xsub{d}{1}$.
Since $[\delta_1,\delta_2]\A{a}{\mu} = \delta_3\A{a}{\mu}$
on solutions of the field equations,
the commutator structure is given by structure constants of the Lie algebra \su2. 
Hence, the gauge symmetry generates a closed group of finite gauge transformations
on solutions $\A{a}{\mu}$.

Associated to the gauge symmetry group are conserved currents
$\duder{\mu} \J{a}{\mu}=0$ given by 
\EQ\label{current}
\J{a}{\mu} = \cross{\mu}{\sigma\nu} \der{\sigma} \K{a}{\nu} 
\endEQ
in terms of the field strengths $\K{a}{\mu}$
for solutions of the field equations.
If we consider two-dimensional hypersurfaces $\Sigma$ of 
the underlying three-dimensional manifold 
on which the fields $\A{a}{\mu}$ are defined,
the flux of the currents $\J{a}{\mu}$ on a given $\Sigma$
defines internal charges carried by the fields on $\Sigma$.
We can evaluate these charges by the surface integral
\EQ
\Q{a} = \int_\Sigma t^\mu \J{a}{\mu}
\endEQ
where $t^\mu$ is a unit normal to $\Sigma$ 
and the integral is understood to use the induced volume element
$t^\mu\vol{\mu\sigma\tau}$ on $\Sigma$.
If $\Sigma$ has topology $R{}^2$, 
then we can express $\Q{a}$ by a line integral
\EQ\label{charge}
\Q{a} = \int_\Sigma t^\mu \J{a}{\mu} = 
\oint_C s^\mu \K{a}{\mu}
\endEQ
where $C$ is the boundary at infinity on $\Sigma$
and $s^\mu$ is the unit tangent to $C$,
in a clockwise orientation.
When there is no current flow across $C$, 
the charges defined by $\Q{a}$ are conserved,
$t^\mu\der{\mu} \Q{a} =0$. 

The charges $\Q{a}$ transform as \su2 vectors 
\EQ
\delta\Q{a} = \cross{bc}{a} \Q{b} \X{c}
\endEQ
under the gauge symmetry \eqrefs{gaugesymm}{Kgaugesymm}
if the functions $\X{c}$ are constant on $C$.
Applying Noether's theorem to this restricted gauge symmetry
yields a direct derivation of the charges $\Q{a}$ 
from the Lagrangian \eqref{L} of the theory. 

A conserved, gauge invariant energy-momentum tensor 
$\duder{\mu} \T{\mu\nu}=0$ 
can be derived from the Lagrangian 
by varying the inverse metric $\invmetric{\mu\nu}$ 
in $L\vol{\sigma\beta\alpha}$ as follows,
\EQ
\T{\mu\nu} = 
\invvol{\sigma\beta\alpha} \delta(L\vol{\sigma\beta\alpha})
/\delta\invmetric{\mu\nu} =
\id{ab}{}( \K{a}{\mu} \K{b}{\nu} - \case{1}{2} \metric{\mu\nu} 
\invmetric{\sigma\tau} \K{a}{\sigma} \K{b}{\tau} )
\endEQ
after some cancellations of terms.
Conservation of $\T{\mu\nu}$ can then be shown 
from the covariance property of the Lagrangian,
$\delta L = \Lie{\zeta}L$
under simultaneous variations of
the inverse metric 
$\delta \invmetric{\mu\nu} = \Lie{\zeta} \invmetric{\mu\nu}$
and the volume form 
$\delta \vol{\sigma\beta\alpha} = \Lie{\zeta} \vol{\sigma\beta\alpha}$
as well as the fields 
$\delta\A{a}{\mu} = \Lie{\zeta} \A{a}{\mu}$
where $\Lie{\zeta}$ is the Lie derivative 
generated by an arbitrary vector field $\vector{\zeta}{\sigma}$.
Gauge invariance of $\T{\mu\nu}$ follows directly from 
the transformation property $\delta\K{a}{\mu} = \cross{bc}{a} \K{b}{\mu} \X{c}$
of the field strengths under the gauge symmetry 
\eqref{gaugesymm} and \eqref{Kgaugesymm}
on solutions of the field equations. 

Conserved currents are obtained from $\T{\mu\nu}$ 
by contracting with a Killing vector field $\kv{\nu}$ of the metric,
yielding the current $\T{\mu\nu}\kv{\nu}$,
where $\duder{(\mu} \kv{\nu)} =0$.
If we consider two-dimensional hypersurfaces $\Sigma$ as above,
these conserved currents then define 
gauge invariant fluxes of energy-momentum and stress
carried by the fields $\A{a}{\mu}$ on a given $\Sigma$,
with $\kv{\nu}$ being respectively a time translation isometry
and a space translation isometry.
(Fluxes of angular momentum are defined 
with $\kv{\nu}$ being a rotation or boost isometry).
The fluxes are given by the surface integral 
\EQ
Q_{\zeta} = \int_\Sigma t^\mu \T{\mu\nu}\kv{\nu}
\endEQ
where $t^\mu$ is a unit normal to $\Sigma$ 
and the integral uses the induced volume element 
$t^\mu\vol{\mu\sigma\tau}$ on $\Sigma$.

\subsection{Gauge Covariant Formulation}

The theory has a natural formulation using 
a field variable $\vecA{\mu}= \A{a}{\mu} \e{a}$
which is an \su2 Lie-algebra valued vector field. 
The \su2 covariant derivative and curvature associated to $\vecA{\mu}$
as a connection geometrically in \su2 \YM/ theory
enter directly into the formulation. 

We start from the \su2 Lie bracket defined by
\EQ
[\vec{\phi},\vec{\psi}] = \cross{bc}{a} \phi^b \psi^c \e{a}
\endEQ
where $\vec{\phi} =\phi^b\e{b}$ and $\vec{\psi} =\psi^b\e{b}$
are arbitrary \su2 Lie-algebra valued fields. 
From the relation \eqref{Bid}
for the auxiliary structure constants $\B{ab}{c}$ 
expressed in terms of the \su2 structure constants $\cross{ab}{c}$,
we then have
\EQs
&& \B{ab}{c} \phi^a \psi^b \e{c} = [[\vec{\phi},\vec{\psi}],\vec{v}]
\\&& \duB{a}{dc} \phi^d \psi^c \e{c} = -[\vec{\phi},[\vec{\psi},\vec{v}]]
\endEQs
which allows all the structures in the theory 
involving $\B{ab}{c}$ and $\duB{a}{bc}$ 
to be formulated using the \su2 Lie bracket $[\ ,\ ]$
and the \su2 vector $\vec{v}= \v{b} \id{}{ba} \e{a}$.

Next we utilize $\vecA{\mu}$ to define the \su2 covariant derivative
\EQ
\Der{\mu} \vec{\phi} = \der{\mu} \vec{\phi} + [\vecA{\mu},\vec{\phi}]
= \der{\mu} \phi^a \e{a} + \cross{bc}{a} \A{b}{\mu} \phi^c \e{a}
\endEQ
The curvature of the covariant derivative is defined through the relation
\EQ
\Der{[\mu}\Der{\nu]} \vec{\phi} =  [\vecF{\mu\nu},\vec{\phi}]
= \cross{bc}{a} \F{b}{\mu\nu} \phi^c \e{a}
\endEQ

To now proceed with the formulation, 
we introduce the field strength $\vecK{\mu}=\K{a}{\mu}\e{a}$ 
for the theory by the algebraic relation
\EQ\label{Krelation}
\vecK{\mu} + \cross{\mu}{\nu\sigma} [\vecK{\nu},[\vecA{\sigma},\vec{v}]]
= \vecF{\mu}
\endEQ
where $\vecF{\mu}=\cross{\mu}{\nu\sigma} \vecF{\nu\sigma}$
is the dual of the \su2 curvature tensor. 
In terms of $\vecK{\mu}$ and $\vecA{\mu}$,
the Lagrangian is given by 
\EQ
L= \invmetric{\mu\nu} (\vecK{\mu}, \vecK{\nu}) + 
\invvol{\sigma\tau\nu} ([\vecK{\sigma},\vecK{\tau}], [\vecA{\nu},\vec{v}])
\endEQ
where $(\ ,\ )$ is the \su2 Killing metric.

The field equations derived from $L$ by varying $\vecA{\mu}$ are given by 
\EQ
2\cross{\mu}{\sigma\tau} ( 
\Der{\sigma}\vecK{\tau} + \case{1}{2} [[\vecK{\sigma},\vecK{\tau}],\vec{v}] ) =0
\endEQ
and the differential identity satisfied by $\vecK{\mu}$ 
for solutions of the field equations is given by 
\EQ 
\invmetric{\sigma\tau} ( \Der{\sigma} \vecK{\tau} 
- [\vecK{\sigma},[\vecK{\tau},\vec{v}]] ) =0
\endEQ

The gauge symmetry on solutions of the field equations is given by 
the infinitesimal transformations
\EQs
&& \delta\vecA{\mu} = \Der{\mu} \vecX + [\vecK{\mu},[\vec{v},\vecX]]
\label{Asymm}
\\&& \delta\vecK{\mu} = [\vecK{\mu},\vecX]
\label{Ksymm}
\endEQs
where $\vecX$ is an \su2 Lie-algebra valued arbitrary function.

Exponentiating the transformations \eqrefs{Asymm}{Ksymm}
generates finite gauge symmetry transformations in the theory as follows. 
First, we express commutators involving $\vecX$ by 
the linear operator 
$\ad{\vecX} = [\ \cdot,\vecX ]$
which acts on \su2 vectors.
We also use the linear operator 
$\proj{\vecX} = -\case{1}{|\vecX|^2} (\ad{\vecX})^2$
which acts as the projection onto \su2 vectors orthogonal to $\vecX$
in the \su2 Killing metric, where $(\vecX,\vecX)= |\vecX|^2$.
Then, calculating $\exp(\delta)$ with the operator $\delta$ given by 
the infinitesimal transformations \eqrefs{Asymm}{Ksymm} 
on $\vecA{\mu}$ and $\vecK{\mu}$
leads to the finite transformations
\EQs
&& \vecA{\mu} \rightarrow 
R_\vecX \vecA{\mu} + R'_\vecX \der{\mu} \vecX 
+ ( R'_\vecX \vecK{\mu}, \vec{v} ) \vecX 
- ( \vecK{\mu}, \vecX ) R'_\vecX \vec{v} 
\label{finiteAsymm}
\\&& \vecK{\mu} \rightarrow 
R_\vecX \vecK{\mu} 
\label{finiteKsymm}
\endEQs
where $R_\vecX = \exp( \ad{\vecX} )$ 
is a rotation generated by $\vecX$ on \su2 vectors,
and $R'_\vecX = \one - \proj{\vecX} 
- \case{1}{|\vecX|^2}\ad{\vecX} ( R_\vecX - \one )$
is a related transformation on \su2 vectors.
Explicitly,
\EQs
&& R_\vecX= 
\one + (\sin |\vecX|) \ad{\vecunitX} - (\one - \cos |\vecX|) \proj{\vecunitX} 
\label{R}
\\&& R'_\vecX = 
\one + \case{1}{|\vecX|} (\one - \cos |\vecX|) \ad{\vecunitX} 
- (\one - \case{1}{|\vecX|} \sin |\vecX|) \proj{\vecunitX} 
\label{R'}
\endEQs
in terms of the unit vector $\vecunitX = \case{1}{|\vecX|} \vecX$.
(The relation between $R_\vecX$ and $R'_\vecX$ is expressed by
\cite{generalref}
the composition formula of rotations 
$R_{\vecX{}_1} R_{\vecX{}_2} = R_{\vecX{}_3}$,
where $\vecX{}_3 = \vecX{}_1 + R'^{-1}_{\vecX{}_1} \vecX{}_2$.)   

The transformations \eqrefs{finiteAsymm}{finiteKsymm}
represent a closed group of finite gauge symmetries
with the novel feature that 
the group acts nonlinearly on $\vecA{\mu}$ but linearly on $\vecK{\mu}$,
where $\vecK{\mu}$ is a nonpolynomial algebraic expression 
in terms of $\vecA{\mu}$ determined by \Eqref{Krelation}.
Algebraically, the group structure of the finite gauge symmetries
is isomorphic to the exponential of 
the \su2 Lie algebra structure \eqref{commutator} 
of the infinitesimal gauge symmetries \eqref{gaugesymm}.

The appearance of $\vec{v}$ 
in the field equations and gauge symmetry of the theory
defines a preferred vector in the \su2 Lie algebra of the gauge group.
Consequently, the theory lacks symmetry invariance 
under rigid \su2 rotations on the field variable
\EQ\label{rotation}
\vecA{\mu} \rightarrow R\vecA{\mu}
\endEQ
where $R=\exp(\ad{\vecX})$ is a constant transformation
generated by an arbitrary \su2 vector $\vecX$, with $\der{\mu}\vecX=0$.

The effect of a transformation \eqref{rotation} in the theory
is to rotate the preferred \su2 vector
\EQ
\vec{v} \rightarrow R \vec{v}
\endEQ
Thus, the direction of $\vec{v}$ can be changed arbitrarily
under field redefinitions given by the transformations \eqref{rotation}.

When $\vec{v}=0$, 
the \su2 symmetry invariance \eqref{rotation} is restored, 
and the theory then reduces to \su2 \YM/ theory.

\section{Extension to other gauge groups}
\label{extension}

We now carry out the extension
from an \su2 gauge group to a general nonabelian gauge group
for the \YM/ theory generalization in \secref{newtheory}.
Since the most general gauge group for \YM/ theory 
is required to be semisimple \cite{generalref},
we only need to consider semisimple gauge groups for the generalization.
The structure of these groups is characterized by 
the Lie algebra of the group being a product of
abelian Lie algebras and nonabelian simple Lie algebras \cite{simple}.

We start with an arbitrary simple Lie algebra
for the gauge group Lie algebra,
with multiplication structure constants $\C{bc}{a}$
and Killing metric components $\k{ab}{} = -\case{1}{2} \C{ad}{e} \C{be}{d}$
in a fixed Lie algebra basis. 
We introduce structure constants $\B{bc}{a}$ 
defining an auxiliary Lie algebra multiplication
which is related to the gauge group Lie algebra multiplication
later through an algebraic condition imposed by gauge invariance. 
Now, setting the number of fields $\A{a}{\mu}$ 
to match the dimension of the gauge group Lie algebra,
we use $\C{bc}{a}$ and $\k{ab}{}$ to replace 
the \su2 structure constants $\cross{bc}{a}$ 
and Killing metric components $\id{ab}{}$ 
in the form of the field strengths $\K{a}{\mu}$,
the Lagrangian $L$, and the gauge symmetry $\delta\A{a}{\mu}$ 
of the \su2 theory given in \Eqsref{K}{gaugesymm}. 

Gauge invariance requires that the Lagrangian $L$ 
vary into a complete divergence under the gauge symmetry $\delta\A{a}{\mu}$.
By the same calculations as followed in the \su2 case,
after using the antisymmetry property $\B{bc}{e} =\B{[bc]}{e}$
and Jacobi property $\B{[bc}{e} \B{d]en}{}=0$ 
of the auxiliary structure constants
in addition to the Jacobi property 
$\C{[bc}{e} \C{d]en}{}=0$ 
and antisymmetry property
$\C{[bc}{e} \k{d]e}{} =\C{bc}{e} \k{de}{}$ 
of the gauge group structure constants,
we find $\delta L$ is a complete divergence up to the term
$\invvol{\nu\sigma\tau} \K{b}{\nu} \K{c}{\sigma} \A{a}{\tau} \X{d} \H{bcad}$
where 
\EQ
\H{bcad}= 
2\C{a[b}{e} \B{|e|c]d}{} - 2\C{d[b}{e}\B{|e|c]a}{} + \C{ad}{e} \B{bce}{} 
\endEQ
with $\B{bce}{} = \B{bc}{a} \k{ae}{}$.
Thus, similarly to the \su2 case, 
we must have
\EQ\label{BCeq}
0=\H{bcad}
\endEQ
which is an algebraic condition relating the structure constants
$\C{bc}{a}$ and $\B{bc}{a}$.

Condition \eqref{BCeq} can be solved,
as shown in \appref{solve},
yielding
\EQ\label{B}
\B{bca}{} = 2 \V{e[b} \C{c]a}{e}
\endEQ
where $\V{eb}=\V{[eb]}$ represent arbitrary constants. 
This expression for $\B{bca}{}$ 
does not automatically satisfy the Jacobi property
$\B{[bc}{n} \B{d]n}{a} =0$
except in the \su2 case as follows.
By the \su2 identities \eqrefs{firstid}{secondid},
$\V{eb}$ can be expressed equivalently as $\cross{eb}{d} \v{d}$
in terms of $\v{d}=\case{1}{2} \cross{d}{eb} \V{eb}$,
so the expression $\B{bca}{} = 2\V{e[b} \cross{c]a}{e}$ 
has the form 
\EQ\label{sutwoB'} 
\B{bca}{}= 2\cross{e[b}{d} \cross{c]a}{e} \v{d} 
= \cross{bc}{e} \cross{ae}{d} \v{d}
\endEQ
The Jacobi property for $\B{bca}{}$ then becomes
a consequence of the \su2 identities,
after some manipulations shown by \Eqrefs{sutwoproperty}{Bid}.

Other than in the \su2 case,
the Jacobi property for $\B{bca}{}$ must be imposed
as an extra condition
\EQs
0= && \B{[bc}{n} \B{d]na}{} = 
\case{4}{3} \k{}{mn} ( \V{e[b} \C{c]n}{e} \V{p[d} \C{m]a}{p} 
+ \V{e[d} \C{b]n}{e} \V{p[c} \C{m]a}{p} 
+ \V{e[c} \C{d]n}{e} \V{p[b} \C{m]a}{p} )
\label{Veq}
\endEQs
which constitutes an algebraic equation on $\V{eb}$.
The solutions of this equation determine 
the form for $\B{bca}{}$ necessary for gauge invariance of 
the \YM/ theory generalization with a general nonabelian simple gauge group. 

Since $\V{eb}$ enters \Eqref{Veq} quadratically,
finding the complete solution is a difficult algebraic problem.
A natural possibility is to consider the expression 
$\V{eb} = \C{eb}{d} \v{d}$ of the same form as works in the \su2 case,
leading to $\B{bca}{} = \C{bc}{e} \C{ae}{d} \v{d}$
similarly to \Eqref{sutwoB'}.
We find the following results for the cases of \su n and \so n gauge groups.
Because of the Lie algebra isomorphism \su2 $\simeq$ \so3,
the \so3 case is the same as the \su2 case. 
When the expression for $\B{bca}{}$ is extended 
from the \su2 case to the \so n case,
condition \eqref{Veq} is satisfied for $n=4$
as a consequence of the Lie algebra isomorphism between \so4 
and a real form of the complexified product \su2 $\times$ \su2.
However, condition \eqref{Veq} fails to be satisfied in any other \so n case. 
The condition also fails to be satisfied 
when the expression for $\B{bca}{}$ is extended to the \su n case 
for any $n\geq 3$.
Hence, among the \su n and \so n cases,
when a relation of the same form 
between the gauge group structure constants and auxiliary structure constants 
as holds in the \su2 case is used,
the \YM/ theory generalization works only in the case of an \so4 gauge group.

We now return to the general case of a nonabelian simple gauge group. 
The simplest alternative possibility to consider 
is the elementary expression $\V{eb} = 2 \u{[e}\v{b]}$
for some constants $\u{e}$ and $\v{b}$.
As shown in \appref{solve},
the condition \eqref{Veq} on $\V{eb}$ is satisfied by having
\EQ\label{V}
\C{dc}{a} \uup{d}\vup{c} =0
\endEQ
where $\uup{d} = \k{}{de}\u{e}$ and $\vup{c}= \k{}{ce}\v{e}$
determine Lie algebra vectors that commute in the gauge group Lie algebra. 
Algebraically, this requires the gauge group Lie algebra to have 
a commutative subalgebra of dimension at least two,
as met by any simple Lie algebra other than \su2. 
Hence, for these simple Lie algebras,
the expression
\EQ\label{Bform}
\B{bca}{} = 2\v{[b} \C{c]a}{e} \u{e} - 2\u{[b} \C{c]a}{e} \v{e} 
\endEQ
determined by \Eqrefs{B}{V} 
satisfies the Jacobi property \eqref{Veq}.
With the relation \eqref{Bform}
between the gauge group structure constants and auxiliary structure constants,
the \YM/ theory generalization works
for all nonabelian simple gauge groups other than \su2. 

The \YM/ theory generalization also works if 
the relation \eqref{Bform} is generalized to 
a sum of similar expressions 
using pairs of Lie algebra vectors that all commute in the simple gauge group Lie algebra.
(In particular, for the case of \su n, 
the number of linearly independent commuting vectors is $n-1$,
and for the case of \so n, the number is $\case{1}{2} n$ if $n$ is even 
and $\case{1}{2} (n-1)$ if $n$ is odd.)
The same relation can used more generally with semi-simple gauge groups,
where the Lie algebra vectors lie in different simple subalgebras 
of the semi-simple gauge group Lie algebra. 

\subsection{Gauge covariant formulation of the extension}

Relation \eqref{Bform} between the gauge group structure constants $\C{bc}{a}$
and the auxiliary structure constants $\B{bc}{a}$
differs from the \su2 case,
giving a somewhat different gauge covariant structure 
for the resulting \YM/ theory generalization.
We now outline this structure. 

We fix Lie algebra basis vectors $\e{a}$
associated to the gauge group structure constants,
\EQ\label{bracket}
[\e{a},\e{b}] = \C{ab}{c} \e{c}
\endEQ
and introduce Lie algebra vectors
$\vec{u} = \uup{a} \e{a}$ and $\vec{v} = \vup{b} \e{b}$ 
defined from the constants $\uup{a}=\id{}{ac}\u{c}$, $\vup{b}=\id{}{bd}\v{d}$.
Using these vectors we define the linear map 
\EQ\label{map}
V(\vec{\phi}) = (\vec{u},\vec{\phi}) \vec{v} - (\vec{v},\vec{\phi}) \vec{u} 
\endEQ
on Lie-algebra vectors $\vec{\phi} = \phi^a \e{a}$.
All structure involving the auxiliary structure constants 
can now be expressed completely in terms of 
the map \eqref{map} and the Lie bracket \eqref{bracket}.

We introduce $\vecA{\mu}=\A{a}{\mu} \e{a}$ 
as the field variable for the gauge covariant formulation of the theory.
We also use the covariant derivative defined in terms of $\vecA{\mu}$ by 
\EQ
\Der{\mu} = \der{\mu} + [\vecA{\mu}, \cdot\ ]
\endEQ
along with the dual of the curvature tensor of this derivative operator,
defined as 
\EQ
\vecF{\sigma} = \cross{\sigma}{\mu\nu} ( 
\der{\mu} \vecA{\nu} + \case{1}{2} [\vecA{\mu},\vecA{\nu}] )
\endEQ

The formulation of the field strengths $\vecK{\mu}= \K{a}{\mu}\e{a}$ 
in the theory is given by the algebraic relation
\EQ
\vecK{\mu} - \cross{\mu}{\nu\sigma} (
V([\vecK{\nu},\vecA{\sigma}]) + [V(\vecK{\nu}),\vecA{\sigma}] )
= \vecF{\mu}
\endEQ
The Lagrangian of the theory has the formulation
\EQ
L= \invmetric{\sigma\tau} ( \vecK{\sigma}, \vecK{\tau} ) + 
2\cross{}{\mu\nu\tau} ( [V(\vecK{\mu}),\vecK{\nu}], \vecA{\tau} )
\endEQ
where $(\ ,\ )$ is the Killing metric in terms of the gauge group structure constants,
such that $(\e{a},\e{b}) = \k{ab}{}$.
The field equations for $\vecA{\mu}$ from varying $L$ have the formulation
\EQ\label{AKfirsteq}
0= 2\cross{\mu}{\sigma\tau} ( 
\Der{\sigma}\vecK{\tau} - [V(\vecK{\sigma}),\vecK{\tau}] )
\endEQ
and the differential identity satisfied by the field strength 
for solutions of the field equations has the formulation
\EQ\label{AKsecondeq}
0= \invmetric{\sigma\tau} ( \Der{\sigma} \vecK{\tau} 
+ V( [\vecK{\sigma},\vecK{\tau}] ) - [V(\vecK{\sigma}),\vecK{\tau}] ) 
\endEQ

The gauge symmetry on solutions of the field equations in the theory 
is given by infinitesimal transformations with the formulation
\EQs
&& \delta\vecA{\mu} = \Der{\mu} \vecX
+ V([\vecK{\mu},\vecX]) - [V(\vecK{\mu}),\vecX]
\\&& \delta\vecK{\mu} = [\vecK{\mu},\vecX]
\endEQs
where $\vecX$ is a Lie-algebra valued arbitrary function.
These transformations have a closed commutator structure
\EQs
&& [\delta_1\vecA{\mu}, \delta_2\vecA{\mu}] = \delta_3\vecA{\mu}
\\&& [\delta_1\vecK{\mu}, \delta_2\vecK{\mu}] = \delta_3\vecK{\mu}
\endEQs
which is the same as the gauge group Lie alebra,
with $\vecX{}_3=[\vecX{}_1,\vecX{}_2]$.
Thus, Lie algebra invariants constructed from $\vecK{\mu}$
yield gauge symmetry invariants in the theory. 

\section{Discussion}
\label{discuss}

In this paper a class of new nonabelian gauge theories
has been constructed for vector fields on three-dimensional manifolds.
These theories describe a generalization of nonabelian \YM/ theory
with a novel nonlinear gauge symmetry and field equations 
for three-dimensional vector potential fields. 

The new theories can be derived by a systematic generalization process
starting from the linear gauge theory of vector potential fields
given by abelian (linearized) \YM/ theory.
The process consists of adding linear and higher order terms 
to the form of the abelian gauge symmetry
while also adding quadratic and higher order terms 
to the form of the abelian field equations
so as to maintain a gauge invariant action principle,
with the condition of gauge invariance
used as an equation to determine the allowed form of terms
added order by order \cite{abelianth,method}.
The linear order terms starting the process are provided by
rigid symmetries special to abelian \YM/ fields $\A{a}{\mu}$ 
in three dimensions
\EQ\label{ymsymm}
\delta\A{a}{\mu} = 
( \C{bc}{a} \A{b}{\mu} 
+ \duB{ab}{c} \k{bd}{} \cross{\mu}{\sigma\nu} \F{d}{\sigma\nu} ) \X{c}
\endEQ
which uses arbitrary constants $\duB{ab}{c} = \duB{[ab]}{c}$,
along with the structure constants $\C{dc}{e}$ 
of the Lie algebra of any nonabelian \YM/ gauge group,
the Killing metric $\k{bd}{}$ of this Lie algebra, 
the cross-product operator $\cross{\mu}{\sigma\nu}$ 
on three-dimensional vectors,
and an arbitrary rigid parameter, $\der{\mu}\X{c}=0$.
For solutions $\A{a}{\mu}$ of the abelian field equations,
where $\F{d}{\sigma\nu}$ is curl of $\A{a}{\mu}$,
the symmetries have a closed commutator structure
$[\delta_1\A{a}{\mu},\delta_2\A{a}{\mu}] = \delta_3\A{a}{\mu}$
which involves $\X{c}$, $\C{ab}{c}$ and $\duB{ab}{c}$.
The existence of these symmetries is limited to three dimensions
because of the dependence on the cross-product operator.
Completing the generalization process 
with the linear terms given by \Eqref{ymsymm}
leads to successively higher order terms,
and fixes $\duB{ab}{c}$ in terms of structure constants 
of an auxiliary Lie algebra related to the Lie algebra of 
the \YM/ gauge group,
producing the striking nonlinearity in the form for
the gauge symmetry and field equations. 

The generalization process can be carried out more broadly
in three dimensions 
starting with the most general form for linear order terms
and requiring the minimum number of derivatives in the form for
higher order terms in the gauge symmetry and field equations.
The outcome of the process leads directly to the new theories,
as can be shown following methods developed in \citeref{method}
This establishes a strong uniqueness result for the theories 
as nonlinear generalizations of abelian \YM/ theory in three dimensions. 

Is there a simple underlying geometrical structure to the theories?
\YM/ theory has a geometrical structure 
which is understood in terms of the vector potential 
as a connection on a fiber bundle. 
In the new theories, the vector potential appears to have
a different geometrical role 
more general than a connection on a fiber bundle,
which is somehow tied to the auxiliary Lie algebra 
related to the gauge group Lie algebra
underlying the structure of the theories. 
Understanding this structure geometrically would be highly worthwhile.

Can the theories be extended to higher dimensions?
The rigid symmetries \eqref{ymsymm} 
needed for the generalization of abelian \YM/ theory in three dimensions
cannot be extended to abelian \YM/ fields in other dimensions. 
However, there are closely analogous symmetries of 
the linear gauge theory of abelian \YM/ fields and antisymmetric tensor fields 
in four (and higher) dimensions,
which can be used to start the generalization process.
The outcome of this process leads to 
a novel nonlinear generalization of four-dimensional \YM/ gauge theory 
with antisymmetric tensor fields. 
A full discussion of this new gauge theory is given in 
a forthcoming paper \cite{newth}.

Investigation of these new gauge theories 
in three and four dimensions
could well hold significant interest for many areas of 
physics and mathematics.

\appendix
\section{Relation between the auxiliary Lie algebra 
and the gauge group Lie algebra}
\label{solve}

We begin by solving condition \eqref{BCeq} 
to obtain the relation \eqref{B} 
for $\B{bc}{a}$ in terms of $\C{bc}{a}$
for any semi-simple gauge group Lie algebra. 
Throughout we raise and lower indices 
on $\B{bc}{a}$ and $\C{bc}{a}$ by $\k{mn}{}$ and its inverse $\k{}{mn}$,
and we freely use the antisymmetry $\B{bca}{}=\B{[bc]a}{}$
and the complete antisymmetry $\C{bca}{}=\C{[bca]}{}$ 
which follows since $\k{mn}{}$ is the Killing metric 
with respect to $\C{bca}{}$.

To proceed, we contract $\k{}{cd}$ onto $\H{bcad}=0$, 
yielding
\EQ\label{trB}
\C{b}{dc} \B{dca}{} = \C{ab}{d} \B{dc}{c}
\endEQ
We next contract $\C{ad}{n}$ onto $\H{bcad}=0$.
After some rearrangements of indices 
and use of the Jacobi property of $\C{bca}{}$, 
we obtain 
\EQ
0=2\B{ncd}{} + \B{cba}{} \C{e}{ba}\C{nd}{e} +\B{eba}{} \C{n}{ba}\C{cd}{e} 
-\case{1}{2} \duB{be}{d} \C{be}{a} \C{anc}{}
\endEQ
Antisymmetrizing on the indices $n,c$ leads to 
\EQ\label{Bterms}
\B{ncd}{} = \case{1}{2} \B{[n|ba}{} \C{e}{ba}\C{|c]d}{e} 
-\case{1}{2} \B{eba}{} \C{[n}{ba}\C{c]d}{e} 
+\case{1}{4} \C{nca}{} \C{be}{a} \duB{be}{d} 
\endEQ
We now rearrange the last term into the form 
$-\case{1}{2} \B{ab}{b} \C{[n|e}{a} \C{c]d}{e}$
using the Jacobi property of $\C{bca}{}$ and the relation \eqref{trB}.
Finally, combining this term with the other terms in \Eqref{Bterms}
yields
\EQ
\B{ncd}{} = \case{1}{2} ( \B{[n|ba}{} \C{e}{ba} -\B{eba}{} \C{[n}{ba} 
- \case{1}{2} \C{[n|e}{a} \B{ab}{b} ) \C{|c]d}{e} 
\endEQ
This relation indicates $\B{ncd}{}$ has the form $2\V{e[n}\C{c]d}{e}$
where $4\V{en}$ is identified with the expression in parenthesis. 

To complete the solution, we show that $\H{bcad}=0$ is satisfied by 
$\B{ncd}{} = 2\V{e[n} \C{c]d}{e}$ with $\V{en}=\V{[en]}$ taken to be arbitrary.
Substituting $\B{ncd}{}$ into $\H{bcad}$ yields
\EQ\label{Cterms}
\H{bcad} = 4 \C{[a|[b}{e} \V{c]n} \C{|d]e}{n} + 
4 \C{[a|[b}{e} \C{c]|d]}{n} \V{ne} + 2 \C{ad}{e} \V{n[c} \C{b]e}{n} 
\endEQ
In \Eqref{Cterms} the middle term vanishes since 
$\C{[a|[b}{e} \C{c]|d]}{n}$ is symmetric in the indices $e,n$
while $\V{ne}$ is antisymmetric.
The first term in \Eqref{Cterms} can be rearranged 
using the Jacobi property of $\C{ab}{e}$ to yield 
$-2\C{ad}{e} \V{n[c} \C{b]e}{n}$
which cancels the last term in \Eqref{Cterms}.
This demonstrates $\H{bcad}=0$. 

We now show that the Jacobi property \eqref{Veq}
on $\B{bcd}{}=2\V{e[b} \C{c]d}{e}$
is satisfied by $\V{eb}= 2\u{[e} \v{b]}$
where $\u{e}$ and $\v{b}$ commute in the gauge group Lie algebra,
$\C{}{ebn} \u{e}\v{b}=0$. 
First, we rearrange \Eqref{Veq} into the form
\EQ
\B{[bc}{n} \B{d]na}{} = 
2\C{na}{p} \C{[c}{ne} \V{b|e|} \V{d]p} 
+ 2\C{a[d}{p} \C{c}{ne} \V{b]e} \V{np} 
\endEQ
Substituting $\V{eb}=2\u{[b}\v{e]}$ leads to 
\EQ
\B{[bc}{n} \B{d]na}{} = 
4\C{na}{p} \C{[c}{ne} \u{b}\v{d]} \u{[e} \v{p]}
+ 2\C{a[d}{p} \C{c}{ne} \u{b]} \u{n} \v{e} \v{p} 
- 2\C{a[d}{p} \C{c}{ne} \v{b]} \u{n} \v{e} \u{p} 
\endEQ
The last two terms directly vanish since $\C{c}{ne} \u{n} \v{e} =0$.
The first term can be rearranged by the Jacobi property
$2\C{c}{n[e} \C{na}{p]} = \C{n}{pe} \C{ac}{n}$,
yielding $2\C{n}{ep} \C{a[c}{n} \u{b}\v{d]} \u{e} \v{p}$
which vanishes since $\C{n}{ep} \u{e} \v{p}=0$.
This demonstrates the Jacobi property $0=\B{[bc}{n} \B{d]na}{}$.
More generally, the same property can be shown to hold if
$\V{eb}$ has the form $2\u{[e}\v{b]} + 2\x{[e} \y{b]} + \cdots$
such that $\u{e},\v{b},\x{e},\y{b},\ldots$ 
all commute in the gauge group Lie algebra.

Finally, for $\V{eb}=2\u{[e}\v{b]}$,
we show that if the commutator of $\u{e}$ and $\v{b}$ 
in the gauge group Lie algebra is nonvanishing,
$\C{}{ebn} \u{e}\v{b} \neq 0$,
the Jacobi property \eqref{Veq} 
on $\B{bcd}{}=2\V{e[b} \C{c]d}{e}$
implies $\u{e}$ and $\v{b}$ belong to an invariant \su2 subalgebra.
Hence, in satisfying the Jacobi property, 
a vanishing commutator of $\u{e}$ and $\v{b}$ 
is a necessary as well as sufficient requirement
in any simple gauge group Lie algebra other than \su2.

To proceed, 
we suppose $\xup{n}=\C{}{ebn} \u{e}\v{b}$ is nonvanishing
and $\k{}{eb} \u{e}\v{b}=0$. 
The Jacobi property \eqref{Veq} then leads to
\EQ\label{xCeq}
0=\xup{e} \C{e[c}{d} \u{b} \v{a]} 
+ \C{[a}{pd} \v{b} \x{c]} \u{p} + \C{[a}{pd} \u{b} \x{c]} \v{p}
\endEQ
First we contract this expression by $\vup{a}\uup{b}\yup{c}$
with an arbitrary $\yup{c}$ satisfying $\u{a}\yup{a} =0 = \v{a}\yup{a}$.
This yields $0=\xup{e}\yup{c} \C{ec}{d}$, 
which implies $\xup{e} \C{ecd}{}$ is proportional to $\u{[c}\v{d]}$.
As a result, we have 
\EQ\label{x}
\xup{e} \C{ecd}{} = 2\u{[c}\v{d]}
\endEQ
The expression \eqref{xCeq} now simplifies to 
$0=\x{[c} ( \C{a}{pd} \v{b]} \u{p} +\C{a}{pd} \u{b]} \v{p} )$.
Contracting by $\xup{c}\uup{b}$ 
and then using the relation \eqref{x} leads to 
\EQ\label{v}
u^2 \v{p} \C{ad}{p} = 2\x{[a}\u{d]}
\endEQ
where $u^2= \k{}{eb} \u{e}\u{b}$. 
Similarly,
\EQ\label{u}
v^2 \u{p} \C{ad}{p} = -2\x{[a}\v{d]}
\endEQ
Taken together, the relations \eqref{x}, \eqref{v}, \eqref{u}
imply $\{\uup{a},\vup{b},\xup{c}\}$
span an invariant \su2 subalgebra in the gauge group Lie algebra.

\section{Structure of the auxiliary Lie algebra}
\label{structure}

The form \eqref{Bform} for the structure constants $\B{bc}{a}$
in the general case of a semi-simple gauge group Lie algebra
yields an auxiliary Lie algebra multiplication
which differs in several features compared to the \su2 case.
In terms of a Lie algebra basis $\e{a}$ 
associated to the structure constants $\C{bc}{a}$,
with multiplication $[\e{b},\e{c}] = \C{bc}{a}\e{a}$
and Killing metric $(\e{b},\e{c})= \k{bc}{}$,
the auxiliary multiplication is given by 
\EQ\label{Bmult'}
[\e{b},\e{c}]_B = 
2(\vec{u},\e{[b}) [\e{c]},\vec{v}] - 2(\vec{v},\e{[b}) [\e{c]},\vec{u}] 
\endEQ
where $\vec{u}= \id{}{ba} \u{b} \e{a}$ 
and $\vec{v}= \id{}{ba} \v{b} \e{a}$ 
are Lie algebra vectors associated to $\u{b}$ and $\v{b}$,
which satisfy $[\vec{u},\vec{v}]=0$.
We can understand the structure of this multiplication
in terms of $\vec{u}$ and $\vec{v}$ as follows.

First, we note the commutator expression \eqref{Bmult'}
depends on $\vec{u}$ and $\vec{v}$ antisymmetrically,
so it remains invariant under 
$\vec{u} \rightarrow a\vec{u} + b\vec{v}$, 
$\vec{v} \rightarrow c\vec{v} + d\vec{u}$
such that $ac-bd=1$.
Using this invariance we set $(\vec{u},\vec{v})=0$ for convenience. 
Next, we let $\Hsp$ define the subspace in the gauge group Lie algebra
such that 
\EQs
&& (\vec{u},\Hsp)=0 =(\vec{v},\Hsp)
\label{uvH}
\\&& [\vec{u},\Hsp]=0=[\vec{v},\Hsp]
\label{uvHcomm}
\endEQs
Thus, vectors in $\Hsp$ are orthogonal to and commute with $\vec{u}$ and $\vec{v}$.
We then let $\Hperp$ define the subspace orthogonal to $\vec{u},\vec{v},\Hsp$
in the gauge group Lie algebra,
\EQs
&& (\vec{u},\Hperp)=0 =(\vec{v},\Hperp)
\label{uvHperp}
\\&& (\Hsp,\Hperp)=0
\label{HHperp}
\endEQs
The spaces $\Hsp$,$\Hperp$, and vectors $\vec{u}$, $\vec{v}$
together span the gauge group Lie algebra.
The property that the gauge group Lie algebra is semi-simple
implies
\EQ\label{uvHperpcomm}
[\vec{u},\Hperp] \subseteq \Hperp, 
[\vec{v},\Hperp] \subseteq \Hperp
\endEQ

Now, from the basis multiplication \eqref{Bmult'},
we have the following auxiliary Lie algebra multiplication.
For vectors $\vec{h},\vec{g}$ in $\Hsp$ and vectors $\vec{x},\vec{y}$ in $\Hperp$,
\EQ
[\vec{u},\vec{h}]_B=0, [\vec{v},\vec{h}]_B=0,
[\vec{h},\vec{g}]_B=0
\endEQ
using the orthogonality \eqref{uvH} 
and commutativity \eqref{uvHcomm},
while 
\EQs
&& [\vec{u},\vec{x}]_B=-|\vec{u}|^2 [\vec{v},\vec{x}],
[\vec{v},\vec{x}]_B=|\vec{v}|^2 [\vec{u},\vec{x}],
\\&& [\vec{x},\vec{y}]_B=0, [\vec{x},\vec{h}]_B=0
\endEQs
using the orthogonality \eqrefs{uvHperp}{HHperp},
where $|\vec{u}|^2=(\vec{u},\vec{u})$
and $|\vec{v}|^2=(\vec{v},\vec{v})$.
In addition,
\EQ
[\vec{u},\vec{v}]_B=0
\endEQ
Algebraically, these commutators show that 
$\vec{u},\vec{v}$ together define an abelian Lie subalgebra $\Asp$,
while $\Hsp$ and $\Hperp$ also define abelian Lie subalgebras,
such that $\Asp$ commutes with $\Hsp$ and acts invariantly on $\Hperp$,
and the subalgebras $\Hsp$, $\Hperp$ commute. 
Hence the span of $\Asp,\Hsp,\Hperp$ defines 
an auxiliary Lie algebra which is the product of $\Hsp$
with the semi-direct product of $\Asp$ and $\Hperp$
where $\frac{1}{|\vec{u}|^2} \vec{u}$ acts on $\Hperp$ by 
multiplication by $-\vec{v}$
while $\frac{1}{|\vec{v}|^2} \vec{v}$ acts on $\Hperp$ by 
multiplication by $\vec{u}$.


\begin{references}
\bibitem{global}
As additional structure, an \su2 bundle over the manifold is needed. 
This bundle is assumed to be globally trivial for simplicity.
\bibitem{isomorphic}
Curiously, this auxiliary Lie alebra is isomorphic to 
the Lie algebra of translations and dilations in Euclidean $R{}^3$ space,
with $\vec{w}{}_1$ and $\vec{w}{}_2$ identified as the generators of translations
and $\case{1}{|\vec{v}|^2} \vec{v}$ identified as the generator of dilations.
\bibitem{simple}
A Lie algebra is simple if and only if it has no proper invariant subalgebras.
\end{references}
\end{document}